\documentclass[12pt]{article}

\usepackage{etex}
\usepackage{amsmath}
\usepackage{amsfonts}
\usepackage{amssymb}
\usepackage{amsthm}
\usepackage{amssymb}
\usepackage[all]{xy}
\usepackage{graphicx}
\usepackage{indentfirst}
\usepackage{diagram}
\usepackage{makecell}
\usepackage[figuresright]{rotating}
\usepackage{pdflscape}
\usepackage{thmtools}
\usepackage{pst-node}
\usepackage{tikz-cd}
\usepackage{setspace}
\usepackage{pgf}        
\usepackage{hyperref}
\usepackage{url}
\usepackage{tikz}
\usetikzlibrary{shapes.geometric, arrows}

\usepackage{epigraph}

\numberwithin{equation}{section}

\makeatletter
\renewcommand{\@biblabel}[1]{#1\hfill \hspace{-0.2cm}}
\makeatother

\usepackage{yfonts}
\newfont{\gothic}{eufm10 at 12pt}

\theoremstyle{plain}
\newtheorem{theorem}{Theorem}[section]
\newtheorem{corollary}[theorem]{Corollary}

\theoremstyle{definition}

\newtheorem{proposition}[theorem]{Proposition}

\numberwithin{equation}{section}
\author{Roman G. Smirnov\footnote{e-mail:  Roman.Smirnov@dal.ca} \\[0.5cm]Department of
  Mathematics and Statistics\\
Dalhousie University\\ Halifax, Nova Scotia, Canada
  B3H~3J5}
\begin{document}
\title{Deriving Production Functions in Economics Through Data-Driven Dynamical Systems}

\maketitle

\begin{abstract}
In their seminal 1928 work, Charles Cobb and Paul Douglas empirically validated the Cobb-Douglas production function through statistical analysis of U.S. economic data from 1899 to 1923. While this established the function's theoretical foundation for growth models like Solow-Swan and its extensions, it simultaneously revealed a fundamental limitation: their methodology could not determine whether alternative production functions might equally explain the observed data.

This paper presents a novel dynamical systems approach to production function estimation. By modeling economic growth trajectories as dynamical systems, we derive production functions as time-independent invariants -- a method that systematically generates all possible functional forms compatible with observed data.

Applying this framework to Cobb and Douglas's original dataset yields two key results: First, we demonstrate that the Cobb-Douglas form emerges naturally from exponential growth dynamics in labor, capital, and production. Second, we show how combining fundamental invariants of this exponential system generates the CES production function as a special case. Our methodology bridges statistical analysis with mathematical systems theory, providing both a verification mechanism for classical results and a tool for discovering new functional relationships.
\end{abstract}

\section{Introduction}
\label{s1} 

Production functions constitute the foundational building blocks of economic modeling, providing the formal mathematical relationship between factor inputs (capital, labor, technology, land, and other productive resources) and economic output (firm-level production, aggregate GDP, and related measures). These functions are typically derived through three complementary approaches: mathematical tractability, empirical calibration, or theoretical consistency. The canonical Cobb-Douglas production function, whose early formalization in economic analysis is attributed to Knut Wicksell, Philip Wicksteed, and L\'{e}on Walras (see \cite{humphrey_1997} for more details and references), gained prominence due to its analytical convenience and theoretically appealing properties --  most notably its  constant returns to scale characteristic, which is an important property in economic contexts. 

In their seminal 1928 paper (\cite{cobb_douglas_1928}; see also \cite{douglas_1976}), Charles Cobb and Paul Douglas used this function that would subsequently  become known as the Cobb-Douglas production function by empirically fitting it to data on U.S. economic growth from 1899 to 1922. Their work represented a pivotal advancement in growth theory, as it was the first to demonstrate that a specific form of the Cobb-Douglas production function could closely match observed patterns in production, capital accumulation, and labor inputs. This empirical validation established the Cobb-Douglas function as a foundational tool for subsequent theoretical and applied work in economics. Specifically, Cobb and Douglas plotted the available data on logarithmic scale, guessed what function could fit well  the data, and then employed statistical tools to verify the accuracy of their assumptions about the form of the production function in question.

This approach, which ultimately legitimized the Cobb-Douglas production function, was considered revolutionary at the time. Unlike in physics -- where fundamental concepts, differential equations, and conserved quantities are typically derived from first principles (e.g., Newton’s second law) -- economic models are often built on heuristic observations, simplifying assumptions, and mathematical convenience. Thus, the methodology employed by Cobb and Douglas in 1928, which relied on rigorous statistical analysis of empirical data, marked a significant breakthrough. It can be summarized as follows:
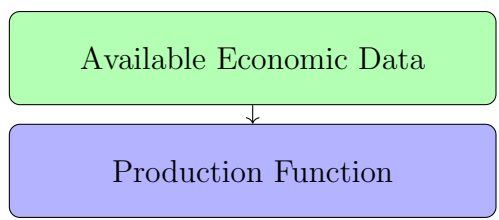
\begin{center}
\tikzstyle{block} = [rectangle, draw, text width=15em, text centered, rounded corners,  minimum height=3em]
\begin{tikzpicture}
[node distance=1.5cm]
\node (n1) at (0,0) [block, fill=green!30]  {Available Economic Data};
\node (n2) [block, below of=n1, fill=blue!30] {Production Function};

\draw [->] (n1) -- (n2);

\end{tikzpicture}
\end{center} 
Despite its significance and its profound influence on the development of economic growth theory over the past century, this approach to deriving production functions leaves many critical questions unanswered. Most notably, even when a production function fits well to a given dataset, key uncertainties remain:
\begin{itemize}
\item Are there alternative production functions that are equally compatible with the data?

\item What is the complete set of production functions that could describe the observed relationships?

\end{itemize} 
These unresolved issues challenge the uniqueness and robustness of such models.

To address these questions, we have developed a novel approach for deriving production functions using dynamical systems theory. In a series of papers (\cite{smirnov_wang_2020, smirnov_wang_2021, smirnov_wang_wang_2022, smirnov_wang_2024}), we have introduced a new framework where production functions emerge as \emph{time-independent invariants} of data-compatible dynamical systems. This method enables the derivation of new production functions that are not empirically derived from economic data \emph{directly}, but rather emerge from dynamical systems determined either by data or mathematical assumptions (\cite{smirnov_wang_2020, smirnov_wang_2024}). The process can be summarized in the following flow diagram:

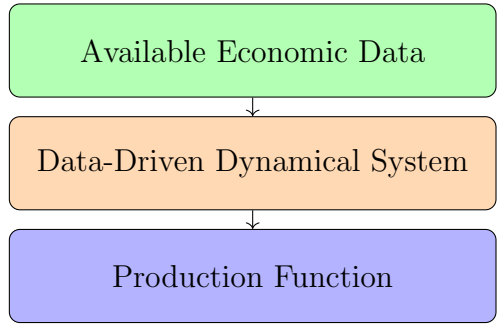
\begin{center}
\tikzstyle{block} = [rectangle, draw, text width=15em, text centered, rounded corners,  minimum height=3em]
\begin{tikzpicture}
[node distance=1.5cm]
\node (n1) at (0,0) [block, fill=green!30]  {Available Economic Data};
\node (n2) [block, below of=n1, fill=orange!30] {Data-Driven Dynamical System};
\node (n3) [block, below of=n2, fill=blue!30] {Production Function};

\draw [->] (n1) -- (n2);
\draw [->] (n2) -- (n3);

\end{tikzpicture}
\end{center} 

In this paper, we review the method and apply it to derive a new production function using the dataset originally analyzed by Charles Cobb and Paul Douglas (\cite{cobb_douglas_1928}). This generalized production function  has the well-known CES (Constant Elasticity of Substitution) production function (\cite{arrow_etal_1961}) as a special case. Furthermore, we establish necessary conditions for a dynamical system driven by exponential growth in labor, capital, and production -- to admit the CES production function as a time-independent invariant.

This paper is organized as follows. Section \ref{s2} presents our novel dynamical systems approach to deriving production functions from economic data, using the exponential growth model that approximates the original Cobb-Douglas dataset \cite{cobb_douglas_1928} as a key example. In Section \ref{s3}, we establish conditions under which the Constant Elasticity of Substitution (CES) production function emerges from this exponential model. Concluding remarks and implications for economic modeling are discussed in Section \ref{s4}.

\section{From data to production functions via dynamical systems} 
\label{s2}

Recall that in their seminal 1928 paper \cite{cobb_douglas_1928}, Charles Cobb and Paul Douglas demonstrated that a production function of the form

\begin{equation}
\label{1}
Y = A L^{\alpha} K^{\beta},
\end{equation}
where $Y=f(L,K)$ represents total production output, $L$ denotes labor input, $K$ --  capital input, $A$ is total factor productivity (a measure of technological efficiency), and $\alpha$ and $\beta$ are the output elasticities of labor and capital, respectively, provided an accurate empirical fit to US economic growth data from 1899 to 1922. Specifically, employing the methods of least squares estimations, the authors assumed constant returns to scale in (\ref{1}), that is $\alpha + \beta = 1$ and obtained a good fit to the data by choosing  $\alpha = 0.75$, which was later verified by the National Bureau of Economic Research to be $0.741$, and determined  $A = 1.01$. The value of $\alpha$ was determined by plotting the time series of production 
($Y$), labor ($L$), and capital ($K$) on a logarithmic scale and analyzing the relative distances between the curves (see \cite{douglas_1976} for details). Recent work by \cite{smirnov_wang_2024} (see also \cite{smirnov_wang_2021} and \cite{smirnov_wang_wang_2022}) has demonstrated that more precise estimates can be obtained through data-driven dynamical systems approaches, which provide a more sophisticated framework for fitting the model to historical data. Specifically, rather than directly fitting the production function in (\ref{1}) to the aforementioned dataset, the authors employed R programming to estimate parameters for the dynamical system: 
\begin{equation}
\label{2}
\dot{L}(t) = b_1 L(t), \quad \dot{K}(t)  =  b_2K(t),  \quad  \dot{Y}(t) = b_3Y(t),
\end{equation} 
which models the exponential growth dynamics of capital, labor, and production. Our analysis, implemented in R, established that the solutions
\begin{equation}
\label{sol}
L(t) = L_0e^{b_1t}, \quad K(t) = K_0e^{b_2t},  \quad Y(t) = Y_0e^{b_3t} 
\end{equation}
were consistent with the empirical results from \cite{cobb_douglas_1928} (cf. \cite{smirnov_wang_2021}) for the parameter values
\begin{equation}
\label{data1}
\begin{array}{lll}

b_1=0.02549605, & \ln L_0=4.66953290 & \mbox{(labor)}, \\

b_2=0.06472564, &  \ln K_0=4.61213588 & \mbox{(capital)},\\ 

b_3=0.03592651, & \ln Y_0 =4.66415363 & \mbox{(production)}.
\end{array}
\end{equation}
Recall, by combining the solutions given by (\ref{sol}) in a way that eliminates time $t$, we produced a one-parameter family of time-independent invariants of the Cobb-Douglas form under a specific linear condition satisfied by the parameters $b_1$, $b_2$, and $b_3$: 
\begin{equation}
\label{inv1}
Y = A L^{\alpha} K^{\frac{b_3}{b_2} - \alpha\frac{b_1}{b_2}}, \quad  0<\alpha<1, 
\end{equation} 
where $\alpha$ is a parameter and $A$ is a constant.

We further showed that the constant-returns-to-scale Cobb-Douglas function ($\alpha + \beta = 1$)
 is a special case of (\ref{inv1}) $b_1<b_3<b_2$.  Notably, the parameter values $b_1$, $b_2$, and $b_3$
from (\ref{data1}), extracted from the dataset in \cite{cobb_douglas_1928}, satisfy this inequality.

Using the bi-Hamiltonian approach introduced in \cite{smirnov_wang_2021}, we obtained explicit formulas for the output elasticities of this particular family member:
\begin{equation}
\label{alphabeta}
\alpha = \frac{b_3 - b_2}{b_1 - b_2}, \quad \beta =   \frac{b_3-b_1}{b_2 - b_1}.
\end{equation}
Using this formula (\ref{alphabeta}) for the parameters $b_1$, $b_2$, and $b_3$ given by (\ref{data1}), we computed the corresponding $\alpha$ and $\beta$: 
\begin{equation}
\label{alphabeta-1}
\alpha=0.7341175376, \quad \beta=0.2658824627.
\end{equation}
Unsurprisingly, these values closely match the original estimates obtained in \cite{cobb_douglas_1928} (see \cite{smirnov_wang_2024} for details).

These findings demonstrate two key insights:

\begin{itemize}
\item[1.] The Cobb-Douglas function (\ref{1}) arises naturally from exponential growth in $K$, $L$, and $Y$
 which may explain its historical empirical fit compared to modern data (cf. \cite{douglas_1976}).

\item[2.] Deriving production functions directly from data-driven dynamical systems provides a more comprehensive framework.
\end{itemize}

In the following section, we extend this approach to CES-like production functions, identifying parameter conditions under which a genuine CES function emerges from the dynamical system (\ref{2}). 

\section{The CES production function revisited} 
\label{s3}

Now, let us employ the techniques of the previous sections to link the Constatn Elasticity of Substitution (CES) production function to the same dynamical system (\ref{2}) shown to be compatible with the data studied in \cite{cobb_douglas_1928}. Recall that the CES production function is given by 
\begin{equation}
\label{ces}
Y = A[\alpha K^p + (1-\alpha)L^p]^{\frac{v}{p}}, 
\end{equation}
where, as before, $Y$ represents production (output); $A$ - the total factor productivity; $\alpha, 0<\alpha<1$ is the share parameter; $K$ and  $L$ are the outputs (capital and labor respectively); $p = \frac{{\sigma}-1}{\sigma}$ is the substitution parameter; $\sigma = \frac{1}{1-p}$ is the elasticity of substitution; $v$ is the degree of homogeneity of the production function (\ref{ces}): if $v = 1$ the function enjoys constant returns to scale, $v<1$ it has  decreasing returns to scale, and $v>1$ indicates it has increasing returns to scale. For a comprehensive analysis of the elasticity of substitution $\sigma$ between labor and capital and its relationship to contemporary economic trends, see \cite{thomas_picketty_2014}.

The production function (\ref{ces}) was originally derived  in \cite{arrow_etal_1961} by fitting a specific linear combination of $\ln K$, $\ln L$, and $\ln Y$ to economic data. In what follows, we will rederive this function from the dynamical system (\ref{2}) under some additional conditions. 

First, we note that from the mathematical viewpoint the formulas (\ref{sol}) may be interpreted as the action of a one-parameter Lie group in the three-dimensional space represented by the economic variables $K$, $L$, and $Y$. Therefore, we should be able to obtain $3$ (the dimension of the space) - $1$ (the dimension of the Lie group acting in the space) $= 2$ fundamental invariants (see \cite{smirnov_wang_2024} and the relevant references therein for more details). The Lie group action is so simple that we can easily derive two fundamental invariants via simple algebra. Indeed, combine first $L(t) = L_0e^{b_1t}$ and  $Y(t) = Y_0e^{b_3t}$. Using the first formula, we first solve for $t$: 
$$t = \frac{1}{b_1}\ln \frac{L}{L_0}.$$ 
Substituting this expression into $Y(t) = Y_0e^{b_3t}$, we arrive at the first fundamental time-independent invariant depending on $Y$ and $K$ given by 
\begin{equation}
\label{i1}
Y = \frac{Y_0}{L_0^{\frac{b_3}{b_1}}}L^{\frac{b_3}{b_1}} \quad (YL^{-\frac{b_3}{b_1}} = \mbox{const}).
\end{equation} 
Combining now the formulas $K(t) = K_0e^{b_2t}$ and $Y(t) = Y_0e^{b_3t}$ and repeating the same procedure by eliminating the parameter $t$, we arrive at the second fundamental time-invariant depending on $Y$ and $L$ given by 
\begin{equation}
\label{i2}
Y = \frac{Y_0}{K_0^{\frac{b_3}{b_2}}}K^{\frac{b_3}{b_2}} \quad (YK^{-\frac{b_3}{b_2}} = \mbox{const}).
\end{equation}

The time-independent functions (\ref{i1}) and (\ref{i2}) are functionally independent and as fundamental invariants have the property that any other time-independent invariant of the Lie group action (\ref{sol}) is a function of (\ref{i1}) and (\ref{i2}). For example, the time-independent invariant (\ref{inv1}) is a function of those. Indeed, it follows from (\ref{i1}) that $$Y = B L^{\frac{b_3}{b_1}},$$ where $B$ is a constant. Hence, $Y^{\alpha} = B^{\alpha}L^{\alpha\frac{b_3}{b_1}}$, $0<\alpha<1$.  On the other hand, it follows from (\ref{i2}) that $$Y = C K^{\frac{b_3}{b_2}},$$ where $C$ is a constant and so $Y^{\alpha} = C^{\alpha}K^{\alpha\frac{b_3}{b_2}}$, $0<\alpha<1$. Therefore, we have 
$$
\begin{array}{rcl}
Y  &= & C K^{\frac{b_3}{b_2}} \\[0.5cm]
&=&  C K^{\frac{b_3}{b_2}} (1)^{\frac{b_1}{b_3}}\\[0.5cm]
& =  & C K^{\frac{b_3}{b_2}} \left(\frac{Y^{\alpha}}{Y^{\alpha}}\right)^{\frac{b_1}{b_3}}\\[0.5cm]
&=& C K^{\frac{b_3}{b_2}} \left(\frac{B^{\alpha}L^{\alpha\frac{b_3}{b_1}}}{C^{\alpha}K^{\alpha\frac{b_3}{b_2}}}\right)^{\frac{b_1}{b_3}}\\[0.5cm]
&=& A L^{\alpha} K^{\frac{b_3}{b_2} - \alpha\frac{b_1}{b_2}}, \quad 0<\alpha<1,
\end{array}
$$
as expected.

To derive the function (\ref{ces}) from the dynamical system (\ref{sol}) determined by the exponential growth in capital ($K$), labor ($L$), and production ($Y$), we will use the  fundamental time-independent invariants (\ref{i1}) and (\ref{i2}) again, but this time we will combine them using addition and subtraction, rather than multiplication and division as in the case of derivation of the family of Cobb-Douglas functions (\ref{inv1}) presented above. Indeed, let us consider the following algebraic expression: 
\begin{equation}
\label{10}
Y^{\frac{1}{b_3}} = \alpha Y^{\frac{1}{b_3}} + Y^{\frac{1}{b_3}} - \alpha Y^{\frac{1}{b_3}},
\end{equation} 
where $ 0<\alpha<1$ is some parameter. Next, using the time-independent invariants (\ref{i1}) and (\ref{i2}) and substituting these expressions instead of the terms of the RHS of (\ref{10}), we obtain
$$
Y^{\frac{1}{b_3}} = \alpha \frac{Y_0^{\frac{1}{b_3}}}{K_0^{\frac{1}{b_2}}}K^{\frac{1}{b_2}} +\frac{Y_0^{\frac{1}{b_3}}}{L_0^{\frac{1}{b_1}}}L^{\frac{1}{b_1}} - \alpha \frac{Y_0^{\frac{1}{b_3}}}{L_0^{\frac{1}{b_1}}}L^{\frac{1}{b_1}} = \alpha \frac{Y_0^{\frac{1}{b_3}}}{K_0^{\frac{1}{b_2}}}K^{\frac{1}{b_2}} + \frac{Y_0^{\frac{1}{b_3}}}{L_0^{\frac{1}{b_1}}}\left(1-\alpha\right)L^{\frac{1}{b_1}},
$$
from which it follows
\begin{equation}
\label{inv2} 
Y = \left[\alpha \frac{Y_0^{\frac{1}{b_3}}}{K_0^{\frac{1}{b_2}}}K^{\frac{1}{b_2}} + \frac{Y_0^{\frac{1}{b_3}}}{L_0^{\frac{1}{b_1}}}\left(1-\alpha\right)L^{\frac{1}{b_1}}\right]^{b_3}, \, 0<\alpha<1.
\end{equation}
Therefore, we conclude that we have proven the following
\begin{proposition}
\label{p1}
The exponential model (\ref{2}) admits the time-independent  family of invariants (\ref{inv2}) that are determined by the fundamental invariants (\ref{i1}) and (\ref{i2}). 
\end{proposition}
We note immediately that the family of functions (\ref{inv2}) looks a lot like the CES production function (\ref{ces}) and under some additional assumptions reduces to it. Indeed, suppose $b_1 = b_2$ in (\ref{2}), that is labor $(L = L(t))$ and capital $(K = K(t))$ enjoy an identical exponential growth. Also we note that without loss of generality we can assume that $L_0 = K_0 = Y_0$. Thus, we can always normalize the data so that the initial condition is equal to $100\%$ -- as it was done in \cite{cobb_douglas_1928}, for example. See also the values of $\ln L_0$ $\ln K_0$, and $\ln Y_0$ given by (\ref{data1}). Therefore, under these assumptions, the formula (\ref{inv2}) reduces to (\ref{ces}) after factoring out the common factor in the RHS of (\ref{ces}) and identifying $\frac{1}{b_1} = \frac{1}{b_2} = p$ and $b_3 = \frac{v}{p}$.  These observations put in evidence that we have the following
\begin{corollary}
\label{c1}
{\em If $b_1 = b_2$ in the exponential model (\ref{2}), it admits the time-independent family of invariants in the form of the CES production function given by (\ref{ces}). }
\end{corollary}
In other words, the CES production function (\ref{ces}), much like the Cobb-Douglas function,  can also be defined as a consequence of the exponential growth in $K$, $L$, and $L$ given by (\ref{2}) under the additional condition that $b_1  = b_2$. 
\section{Conclusions} 
\label{s4}

The Cobb-Douglas production function (\ref{1}) was empirically validated by Cobb and Douglas (1928) using U.S. economic data from 1899–1922. However, their approach left open questions about the uniqueness and exhaustiveness of this functional form. In this paper we have briefly reviewed the fact that 
 the Cobb-Douglas function emerges naturally as a time-independent invariant of a dynamical system modeling exponential growth in   labor ($L$), capital ($K$), and the production ($Y$).   We have shown that the system’s parameters $(b_1, b_2, b_3)$ derived from the dataset originally studied by Cobb and Douglas directly determine the elasticities $\alpha$ and $\beta$ in (\ref{1}) via the formula (\ref{alphabeta}), yielding the values (\ref{alphabeta-1}) that closely match the historical estimates obtained in \cite{cobb_douglas_1928}. The paper discusses a data-driven dynamical systems approach introduced recently to derive production functions, moving beyond ad hoc fitting. Furthermore, we have shown that the exponential model (\ref{2}) compatible with the dataset studied in \cite{cobb_douglas_1928} admits the one-parameter family of Cobb-Douglas production functions (\ref{inv1}). Moreover, we have rederived this family of functions as a combination of the two fundamental invariants (\ref{i1}) and (\ref{i2}) associated with the group action (\ref{sol}). This method generalizes to other production functions (e.g., CES (\ref{ces})) and provides a rigorous mathematical foundation for empirical calibration. Specifically, the Constant Elasticity of Substitution (CES) function (\ref{ces}) is  derived in this paper as another invariant of the same dynamical system, but under the condition $b_1 = b_2$ by introducing a recombination of the fundamental invariants (\ref{i1}) and (\ref{i2}). This result links CES parameters ($v$, $p$) to the dynamical system’s growth rates (Proposition \ref{p1} and Corollary \ref{c1}), offering new insights into its theoretical validity and empirical applicability.

\section*{Acknowledgments }

The author thanks the organizers of the 43rd International Conference on Mathematical Methods in Economics (MME 2025) for the invitation to attend the conference and give a lecture.

\end{document}